\documentclass[aps,prd,amssymb,amsfonts,twocolumn,nofootinbib]{revtex4}
\usepackage{amsmath,amssymb}
\usepackage[dvips]{graphicx}
\usepackage{url}

\begin{document}

\title{Primordial magnetic field from non-inflationary cosmic expansion\\
in Ho\v{r}ava-Lifshitz gravity}

\author{Satoshi Maeda$\,^{\rm a}$, Shinji Mukohyama$\,^{\rm b}$
and Tetsuya Shiromizu$\,^{\rm a}$}

\affiliation{$^{\rm a}$Department of Physics, Kyoto University, Kyoto 606-8502, Japan}

\affiliation{$^{\rm b}$IPMU, The University of Tokyo, Kashiwa, Chiba 277-8582, Japan}

\date{\today}
\begin{abstract}
 The origin of large-scale magnetic field in the universe is one of the
 greatest mysteries in modern cosmology. We present a new mechanism for 
 generation of large-scale magnetic field, based on  the power-counting
 renormalizable theory of gravitation recently proposed by
 Ho\v{r}ava. Contrary to the usual case in general relativity, the
 $U(1)$ gauge symmetry of a Maxwell action in this theory permits 
 terms breaking conformal invariance in the ultraviolet. Moreover, for
 high frequency modes, the anisotropic scaling intrinsic to the theory
 inevitably makes the sound horizon far outside the Hubble 
 horizon. Consequently, non-inflationary cosmic expansion in the early
 universe naturally generates super-horizon quantum fluctuations of the
 magnetic field. Specializing our consideration to the case with the 
 dynamical critical exponent $z=3$, we show an explicit set of
 parameters for which (i) the amplitude of generated magnetic field is
 large enough as a seed for the dynamo mechanism; (ii) backreaction to
 the cosmic expansion is small enough; and (iii) the high-energy
 dispersion relation is consistent with the most recent observational
 limits from MAGIC and FERMI. 
\end{abstract}
\maketitle

\section{Introduction}
\label{sec:intro}

Ho\v{r}ava recently proposed a class of power-counting renormalizable
theories of gravity~\cite{Horava:2009uw}. The power-counting
(super-)renormalizability stems from the Lifshitz-type anisotropic
scaling 
\begin{equation}
 t\rightarrow b^zt,~~
  \vec{x}\rightarrow b\vec{x},
\label{scale}
\end{equation}
with the dynamical critical exponent $z=3$ (or $z>3$). Because of this
scaling, the theory is often called Ho\v{r}ava-Lifshitz gravity. Although 
renormalizability of matter action, e.g. the standard model action, does
not require the anisotropic scaling, quantum corrections should generate
terms leading to the anisotropic scaling with a common $z$ for all
physical degrees of freedom in the ultraviolet (UV). For these reasons,
in the present paper we shall seriously consider the anisotropic scaling
with $z\geq 3$ for matter degrees of freedom, especially for photon.

Note that the value of $z$ ($\geq 3$) in the UV is a part of the
definition of a theory, provided that the theory is renormalizable. Once
$z$ in the UV is fixed then terms leading to higher $z$ would not be
generated by quantum corrections. In this paper, for simplicity, we
shall restrict our consideration to the simplest case where $z$ in the
UV is $3$. However, in principle the mechanism presented in this paper
works for any values of $z$ in the UV. (See the last paragraph of
Sec.~\ref{sec:summary} for a comment on the case with general $z>3$.)

Cosmology based on this theory has been investigated by many authors,
and number of interesting implications have been pointed 
out~\cite{Calcagni:2009ar,Kiritsis:2009sh,Mukohyama:2009gg,Yamamoto:2009tf,Mukohyama:2009mz,Saridakis:2009bv}.
Among them, the one particularly relevant to the present paper is that
the anisotropic scaling of a physical degree of freedom leads to a
new mechanism for generation of super-horizon quantum fluctuations
without inflation~\cite{Mukohyama:2009gg}.

Needless to say, the driving force behind recent enthusiasm for
cosmology based on Ho\v{r}ava-Lifshitz gravity is the fact that this
theory is a new candidate for quantum gravity. At this moment, it is not
yet clear if some version of Ho\v{r}ava-Lifshitz gravity makes sense at
the quantum level and can be applied to the real world. Indeed, the
version without the projectability condition is already known to be
problematic~\cite{Charmousis:2009tc}. On the other hand, problems
pointed out in the literature are absent if the projectability condition
is maintained and if the detailed balance condition is
abandoned~\cite{Mukohyama:2009mz}. Therefore, Ho\v{r}ava-Lifshitz 
gravity with the projectability condition without the detailed balance
condition has a potential to be theoretically consistent and
phenomenologically viable. While there still remain many issues to be
addressed in the future, we may regard Ho\v{r}ava-Lifshitz theory as a
candidate for the UV completion of general relativity. For this reason,
it is interesting and important to investigate cosmological implications
of the theory.

In this paper we shall focus on the origin of large-scale magnetic field
in the universe. The observed magnitude of the magnetic field at scales
of galaxies and clusters is about $1\mu$Gauss. At larger scales, on the
other hand, there is only an upper limit ($<10^{-9}$Gauss) from 
e.g. observation of the cosmic microwave background~\cite{CMB}. Galactic
magnetic field can be amplified by the so-called dynamo
mechanism~\cite{Grasso:2000wj}, but a seed magnetic 
field must be provided by other mechanisms in an earlier epoch since
the dynamo mechanism does not generate magnetic field from nothing. As
for scales of clusters, efficient amplification mechanism from tiny
primordial magnetic fields to the observed amplitudes has not been
established. In this sense there is no consensus about how observed  
magnetic fields at cluster scales could be related to primordial
ones. However, it has been suggested that, once galactic magnetic field
is amplified by the dynamo mechanism, those amplified magnetic fields
can spread over cluster scales through galactic outflows
(winds/active galactic nuclei ejecta)~\cite{ICM1} or some plasma
instabilities~\cite{ICM2}. For these reasons, in the present paper, we
shall restrict our consideration to the origin of primordial magnetic
seed fields responsible for galactic magnetic fields.

Various generation mechanisms of the seed magnetic field have been
proposed so far. Among them, scenarios based on cosmic inflation are 
popular~\cite{Turner:1987bw,Bamba:2003av}. (See, however,
Refs.~\cite{Grasso:2000wj,Kahniashvili:2009qi,Ichiki1,defects,Lemoine:1995vj} 
for other scenarios.) This is largely because inflation provides a
natural framework in which physical wavelengths of quantum fluctuations 
are stretched by rapid expansion to scales beyond the Hubble
horizon. However, this (quasi-)standard paradigm based on inflation was
recently challenged on the basis of strong
backreaction and strong coupling~\cite{Demozzi:2009fu}. 
If we simply demand that the generated magnetic field does not
backreacts to the dynamics of  inflation significantly and that the
coupling constant in the theory is not extremely large then the
amplitude of the primordial seeds cannot exceed $10^{-32}$G in Mpc
scales. Since a large coupling constant makes any perturbative
calculations untrustable, it is necessary to take the strong
backreaction into account or/and to analyze the strongly-coupled quantum
dynamics non-perturbatively. In some model of inflationary
magnetogenesis, the effect of backreaction was investigated and it was
claimed that the magnetic field generated by inflation significantly
backreacts to the cosmic expansion in such a way that generation of
magnetic field is, after all, highly suppressed. At present it is not
clear whether the strong backreaction or/and the strong coupling really
spoils other models of inflationary magnetogenesis or not. 
While it is certainly worthwhile investigating this issue in more
details, it is also plausible to look for alternative mechanisms.

It is well known that conformal invariance of the standard Maxwell
action prevents cosmic expansion (including inflation) from acting as a
generation mechanism of magnetic field \cite{Parker:1968mv}: the Maxwell field in the flat
Friedmann-Robertson-Walker universe does not feel cosmic expansion and
behaves as if it were in flat spacetime. Therefore, any generation
mechanisms need to include, one way or another, effects breaking
conformal invariance.

Interestingly enough, in Ho\v{r}ava-Lifshitz gravity, the $U(1)$ gauge
symmetry of a Maxwell action permits terms breaking conformal
invariance. Actually, among them, most important in the UV are those
associated with the anisotropic scaling (\ref{scale}). Therefore,
breaking of conformal invariance is not only possible but also
inevitable in the UV regime of Ho\v{r}ava-Lifshitz gravity. Moreover, as
already stated, super-horizon quantum fluctuations can be generated
without inflation~\cite{Mukohyama:2009gg}. The essential reason is that
the sound horizon for high frequency modes is far outside the Hubble
horizon if a physical degree of freedom exhibits the anisotropic
scaling.

The rest of this paper is organized as follows. In Sec.~\ref{sec:action}
we describe the action for an electromagnetic field in
Ho\v{r}ava-Lifshitz theory. In Sec.~\ref{sec:generation} we describe our
mechanism for generation of magnetic fields and present the power
spectrum of the magnetic field. The results obtained by qualitative
scaling arguments there will be confirmed by explicit calculations in
Appendix. In Sec.~\ref{sec:backreaction} we shall investigate the
backreaction problem raised in Ref. \cite{Demozzi:2009fu} and confirm that
the backreaction is small enough for a wide range of parameters. Then we
shall estimate the order of magnitude of the generated magnetic field in
Sec.~\ref{sec:evaluation}. We shall show an explicit set of parameters
for which (i) the amplitude of generated magnetic field is large enough
as a seed for the dynamo mechanism; (ii) backreaction to the cosmic
expansion is small enough; and (iii) the high-energy dispersion relation
is consistent with the most recent observational limits from MAGIC
\cite{Albert:2007qk} and FERMI \cite{Collaborations:2009zq}. Finally,
Sec.~\ref{sec:summary} is devoted to a summary of this paper.

\section{electromagnetic field in Ho{\v r}ava-Lifshitz theory}
\label{sec:action}

In Ho\v{r}ava-Lifshitz theory, gravity is described by three basic
quantities: the lapse function $N(t)$, the shift vector $N^i(t,\vec{x})$
and the three-dimensional spatial metric $g_{ij}(t,\vec{x})$. We can
combine these three to form a $4$-dimensional metric of the ADM form: 
\begin{equation}
 ds^2 = -N^2dt^2 + g_{ij}(dx^i+N^idt)(dx^j+N^jdt). 
\end{equation}
The fundamental symmetry of the theory is invariance under the
foliation-preserving diffeomorphism:
\begin{equation}
 t\to t'(t), \quad \vec{x}\to \vec{x}'(t,\vec{x}). 
  \label{eqn:symmetry}
\end{equation}
This symmetry, combined with the value of the dynamical critical
exponent $z$ ($\geq 3$) in the UV, completely determines the structure
of the gravitational action~\cite{Horava:2009uw,Sotiriou:2009gy}.

In this paper we investigate the dynamics of the electromagnetic field,
i.e. a $U(1)$ gauge field, in Ho\v{r}ava-Lifshitz theory. The basic
quantities are the scalar potential $A_0(t,\vec{x})$ and the vector
potential $A_i(t,\vec{x})$. The (generalized) Maxwell action must
respect the $U(1)$ gauge symmetry as well as the foliation-preserving
diffeomorphism invariance. As in the gravity sector, these symmetries,
combined with the value of the dynamical critical exponent $z$ ($\geq
3$) in the UV, completely determine the structure of the action. The
(generalized) Maxwell action is, thus, 
\begin{eqnarray}
 S = \frac{1}{4}\int N\sqrt{g}dtd^3\vec{x}
   \Bigg[\frac{2}{N^2}g^{ij}(F_{0i}-N^kF_{ki})
   \nonumber\\
 \times (F_{0j}-N^lF_{lj}) - G[B_i]\Bigg],
 \label{eqn:generaliezdMaxwell}
\end{eqnarray}
where $F_{0i}=\partial_0 A_i- \partial_i A_0$,
$F_{ij}=\partial_iA_j-\partial_jA_i$, and $G[B_i]$ is a function of 
the magnetic field $B_i$ and its spatial derivatives. The magnetic field
is defined, as usual, by 
\begin{eqnarray}
 B_i = \frac{1}{2}\epsilon_{ijk}g^{jl}g^{km}F_{lm},
\end{eqnarray}
where $\epsilon_{ijk}$ is the totally anti-symmetric tensor with
$\epsilon_{123}=\sqrt{g}$. Restricting our consideration to the case
where $A_i$ is a free field, $G[B_i]$ in general has the form 
\begin{eqnarray}
G[B_i] & = & a_1B_iB^i 
 +a_2g^{ik}g^{jl}\nabla_iB_j\nabla_kB_l \nonumber \\
& &  +a_3g^{il}g^{jm}g^{kn}\nabla_i\nabla_jB_k\nabla_l\nabla_mB_n
 + \cdots,
\label{potential}
\end{eqnarray}
where $a_1$, $a_2$ and $a_3$ are constants and $\nabla_i$ is the spatial
covariant derivative compatible with $g_{ij}$. The highest derivative
term in $G[B_i]$ is the square of the ($z-1$)-th derivative of the
magnetic field. The (generalized) Maxwell action
(\ref{eqn:generaliezdMaxwell}) is a special case of the vector field
action considered in Ref. \cite{Kiritsis:2009sh}.

It is easy to see that the scaling dimension of $A_0$ and $A_i$ are
$(z+1)/2$ and $(3-z)/2$, respectively. This will be important for the
estimate of the power-spectrum of the magnetic field in the next
section.

One of most important properties of Ho\v{r}ava-Lifshitz theory is that
in the UV the theory exhibits the anisotropic scaling with the dynamical
critical exponent $z=3$~\footnote{Theories with the dynamical critical
exponent larger than $3$ are power-counting super-renormalizable and,
thus, worthwhile considering. In this paper, for simplicity, we consider
the case with $z=3$ only.}. As already stated in introduction, this
property should be shared with matter fields such as the electromagnetic
field. Therefore, the function $G[B_i]$ in the UV should be dominated by
the $z=3$ term 
\begin{eqnarray}
G[B_i] \ni
 \frac{1}{M^4}
 g^{il}g^{jm}g^{kn}\nabla_i\nabla_jB_k\nabla_l\nabla_mB_n,
\end{eqnarray}
where $M$ is a mass scale defined by $a_3=1/M^4$. From the stability of
the system in the UV, the sign of this term is required to be
positive. For lower energy scales, relevant deformations, i.e. terms
with less number of spatial derivatives, become important.

Before closing this section, let us briefly mention observational bounds
on $M$. The electromagnetic field in our model has dispersion 
relation~\footnote{Rigorously speaking, each coefficient on the right
hand side is subject to logarithmic running under renormalization group
flow. How they actually run has not yet been investigated in full
details. (See, however, Ref. \cite{Iengo:2009ix} for analysis in a simplified
setup.) Among them, the running of the coefficient of $k_{phys}^2$ is
required to be small. In the present paper, we suppose that this is
already achieved by tuning various coupling constants. As for other
coefficients, argument in the present paper does not depend on their
precise values and we can treat them as constant unless they change by
many orders of magnitude.} 
\begin{eqnarray}
 \omega^2 \simeq \frac{k_{phys}^6}{M^4}
  + \kappa\frac{k_{phys}^4}{M^2} + k_{phys}^2,
\end{eqnarray}
where $k_{phys}$  is the physical wavenumber and $\kappa$ is defined by
$a_2=\kappa/M^2$. This leads to the energy-dependent photon velocity 
\begin{eqnarray}
 v & = & \frac{d\omega}{dk_{phys}} = \frac{k_{phys}}{\omega}
  \left(1+2\kappa\frac{k_{phys}^2}{M^2}+3\frac{k_{phys}^4}{M^4}\right)
  \nonumber\\
 & \simeq &
  1 + \frac{3}{2}\kappa\frac{k_{phys}^2}{M^2}
  + \left(\frac{5}{2}-\frac{5}{8}\kappa^2\right)\frac{k_{phys}^4}{M^4}
  + O\left(\frac{k_{phys}^6}{M^6}\right). 
\end{eqnarray}
For $\kappa=O(1)$, the leading correction to the photon velocity comes
from the term proportional to $k_{phys}^2/M^2$. In this case the MAGIC
Collaboration~\cite{Albert:2007qk} and the Fermi GBM/LAT
Collaborations~\cite{Collaborations:2009zq} give similar lower bound on
$M$: 
\begin{eqnarray}
 M > 10^{11}GeV. \label{eqn:boundM}
\end{eqnarray}
If $|\kappa|\ll 1$ then the leading correction comes from the term
proportional to $k_{phys}^4/M^4$ and the lower bound on $M$ is weaker.

\section{Generation of super-horizon scale magnetic field without
 inflation}
\label{sec:generation}

Let us consider the electromagnetic field described by the (generalized)
Maxwell action (\ref{eqn:generaliezdMaxwell}) in the flat
Friedmann-Robertson-Walker (FRW) spacetime. The metric is given by 
\begin{eqnarray}
 ds^2 & = & -dt^2+g_{ij}dx^i dx^j \nonumber \\
      & = & -dt^2 + a^2\delta_{ij} dx^idx^j \nonumber \\
      & = & a^2[-d\eta^2+\delta_{ij}dx^i dx^j ],
\label{metric}
\end{eqnarray}
where $a$ is the scale factor and $\eta$ is the conformal 
time. The Latin indices run over spatial index ($i=1,2,3$).

Let us first look at the freeze-out condition by assuming the power-law
expansion for the background universe~\cite{Mukohyama:2009gg}.
As seen easily from the action, the dispersion relation for 
the vector potential will be significantly modified as
\begin{equation}
\omega \sim k^z/a^zM^{z-1},
\end{equation}
where $\omega$ is the physical frequency and $k$ is the comoving wavenumber. 
In the UV the dynamical critical exponent $z$ is $3$ but we shall leave
it as a free parameter for a while until we need to specify
it. Hereafter, for simplicity we adopt the unit with $M=1$. The
fluctuation is expected to oscillate (freeze out) if $\omega \gg H$
($\omega \ll H$). Thus if 
\begin{eqnarray}
\partial_t\left(a^{2z}H^2\right)>0.
\label{freeze-out}
\end{eqnarray}
is satisfied, vector fields first oscillate and then freeze out 
afterwards. For the power-law expansion, $a\propto t^p$, 
the condition of Eq. (\ref{freeze-out}) becomes $p>1/z$. 
Throughout this paper we consider cases satisfying the above condition
since in this case super-horizon quantum fluctuations of vector fields
can be generated without inflation. 


Now we compute the power spectrum of the magnetic field in a qualitative way
(See Appendix for quantitative analysis). 
It is easy to guess the scale dependence of it 
by using the scaling dimension of the fields. 
The vector potential $A_i$ has the kinetic term of 
the form
\begin{equation}
 \frac{1}{2}\int d\eta d^3\vec{x}
  (\partial_{\eta}A_i)^2.
  \label{eqn:kinetic-term}
\end{equation}
Here remember that a canonically normalized scalar field 
has the kinetic term of the form 
$\frac{1}{2}\int d\eta d^3\vec{x}a^2(\partial_{\eta}\phi)^2$
\cite{Calcagni:2009ar,Mukohyama:2009gg}. 
Thus, the behavior of $\tilde{A}_i\equiv A_i/a$ 
should be similar to a canonically normalized scalar field. 
For simplicity, we consider a power-law expansion 
\begin{equation}
 a \sim \alpha\eta^q \propto t^p, 
\end{equation}
where we see from the definition of the conformal time that 
$ t\sim \alpha\eta^{1+q}$, $p=q/(1+q)$ and $q=p/(1-p)$. For this background, 
the ratio of $H$ to $\omega$ becomes 
\begin{equation}
 \frac{H}{\omega} \sim \frac{Ha^z}{k^z} \sim \alpha^{z-1}k^{-z}\eta^{qz-q-1}. \label{ratio}
\end{equation}

When $H\sim\omega$, all relevant time scales agree. Thus, at
the {\it sound}-horizon crossing $H\sim\omega$, the
power-spectrum of the ``canonically normalized'' field $\tilde{A}_i$ 
should follow from the scaling dimension of $\tilde{A}_i$ as
\begin{eqnarray}
 \left.{\cal P}_{\tilde{A}_i}\right|_{H\sim\omega}
  & \sim &    \left. H^{(3-z)/z}\right|_{H\sim\omega} \nonumber \\
  & \sim &  \alpha^{-(3-z)/z}\left. \eta^{-(q+1)(3-z)/z}\right|_{H\sim\omega} \nonumber \\
  & \sim &  \left[\alpha k^{-(q+1)}\right]^{(3-z)/(qz-q-1)}, 
  \label{eqn:P-horizon-crossing}
\end{eqnarray}
where we used $k^z \sim \alpha^{z-1}\eta^{qz-q-1}$ which holds for $H \sim \omega$. 

From the kinetic term of Eq.(\ref{eqn:kinetic-term}), it is easy to see that
the super-horizon growing mode behaves as 
\begin{equation}
 \left.A_i\right|_{H\gg\omega} \propto \eta. 
\end{equation}
Thus, at the super-horizon scale, the time evolution of ${\cal P}_{\tilde{A}_i}$ 
is given by 
\begin{equation}
 \left.{\cal P}_{\tilde{A}_i}\right|_{H\gg\omega}
  \propto \frac{\eta^2}{a^2} \propto \eta^{2(1-q)}
  \propto \left(\frac{H}{\omega}\right)^{2(1-q)/(qz-q-1)},
  \label{eqn:P-evolution}
\end{equation}
where we used Eq. (\ref{ratio}) in the last. For the moment, we did not 
take care of the wavenumber dependence. Recovering the wavenumber dependence, 
then, we obtain 
\begin{eqnarray}
 \left.{\cal P}_{\tilde{A}_i}\right|_{H\gg\omega}
  & \sim & 
  \left.{\cal P}_{\tilde{A}_i}\right|_{H\sim\omega} 
  \times \left(\frac{H}{\omega}\right)^{2(1-q)/(qz-q-1)} \nonumber \\
  & \sim & \alpha^mk^{n-2}\frac{\eta^2}{a^2},
\end{eqnarray}
where 
\begin{equation}
  m := \frac{z-1}{qz-q-1}, \quad
   n := 5 - \frac{z}{qz-q-1}. 
\end{equation}
Thus, we compute $\left.{\cal P}_{A_i}\right|_{H\gg\omega}$ as 
\begin{equation}
 \left.{\cal P}_{A_i}\right|_{H\gg\omega}
  \sim \alpha^mk^{n-2}\eta^2, 
\end{equation}
and the power spectrum of the magnetic field is 
\begin{equation}
 {\cal P}_{B} \sim \frac{k^2}{a^4}{\cal P}_{A_i}
  \sim \alpha^mk^n\frac{\eta^2}{a^4}. 
\end{equation}
So far, we have been working in the unit with $M=1$. Noting that 
${\cal P}_{B}$ has mass dimension four, we can easily recover $M$ as 
\begin{equation}
 {\cal P}_{B} 
  \sim (\alpha M)^mk^n\frac{\eta^2}{a^4}. \label{sp}
\end{equation}
(See Appendix for explicit confirmation of this result.)

The correlation length of the generated magnetic fields is roughly the
sound-horizon size. Since the anisotropic scaling in the UV regime makes
the sound horizon far outside the Hubble horizon, the magnetic field on
super-horizon scales can be generated. Hence the correlation length
naturally becomes the cosmological scale. 

We find that time evolution of the power spectrum of the super-horizon
magnetic field is proportional to $ \eta^2/a^4$. Here if $a$ is the de
Sitter expansion $a\propto 1/\eta$, then we obtain 
$\mathcal{P}_{B} \propto a^{-6}$.
Thus, the generated magnetic field rapidly decays. 
In this sense inflationary universe is not good for the generation of
the magnetic field in our model. Therefore we will not 
consider inflationary phases.


\section{Absence of backreaction problem}
\label{sec:backreaction}

As recently pointed out in Ref. \cite{Demozzi:2009fu}, we have to check 
if the generated magnetic fields affect the background universe. Let us
suppose that the $z=3$ regime, where magnetic fields at super-horizon
scales are generated, begins at $\eta=\eta_{\rm s}$ and ends at
$\eta=\eta_{\rm out}$. Here, $\eta_{\rm out}$ is determined by 
$H(\eta_{\rm out})=M$ and we shall take the limit 
$\eta_{\rm s}\to -\infty$ in the end of calculation. Then, the total
energy density of magnetic fields for $\eta>\eta_{\rm out}$ is 
\begin{eqnarray}
\epsilon_B(\eta) \simeq \int_{k_f(\eta_{\rm s})}^{k_f(\eta_{\rm out})}{\cal P}_B \frac{dk}{k},
\end{eqnarray}
where $k_f(\eta_{\rm out})$ and $k_f(\eta_{\rm s})$ stand for the
wavenumbers of fluctuations which freeze out at $\eta_{\rm out}$ and
$\eta_{\rm s}$, respectively. Here, for simplicity, we have assumed that
$\kappa=O(1)$ or $|\kappa|\ll 1$ so that the $z=2$ regime is short or
absent. Then we see that $n=2(5q-4)/(2q-1)>0$ is necessary and
sufficient for the finiteness of the integral in the limit 
$\eta_s\to -\infty$. Next, one wonders if electric fields affect the
background universe. Since ${\cal P}_E \propto {\cal P}_B/k^2$, we
realize that $n-2>0$ is necessary and sufficient for the finiteness of
the total energy density of electric fields. 

Together with the freeze-out condition of Eq. (\ref{freeze-out}), we have the 
constraint for the power of the expansion rate as $q>1$ ($1/2<p<1$). 

At $\eta=\eta_{\rm out}$, $\epsilon_B \sim M^4$ and the background
energy density is around $H^2M_{\rm pl}^2\sim M^2M_{\rm pl}^2$, where
$M_{\rm pl} \sim 10^{19}{\rm GeV}$ is the Planck scale. Then the
backreaction from the generated magnetic field will be negligible if 
$M \ll M_{\rm pl}$. Importantly, this is compatible with the lower bound
(\ref{eqn:boundM}).

In summary, the backreaction problem does not appear in the cases with
$q>1$ ($1/2<p<1$) and $M \ll M_{\rm pl}$.

\section{The evaluation of the generated magnetic field at equal time}
\label{sec:evaluation}

In order to estimate the magnitude of generated magnetic fields, we need
to specify the FRW background evolution in the early universe. For
simplicity, we assume that an oscillating scalar field dominates the
evolution of the background FRW universe in the early stage and then
reheats the universe at $\eta=\eta_{\rm rh}$. Thus, we set $q=2$ for
$\eta\leq\eta_{\rm rh}$. We also suppose that 
$\eta_{\rm out}<\eta_{\rm rh}<\eta_{\rm eq}$, where 
$\eta=\eta_{\rm eq}$ corresponds to the matter-radiation equality. Then
the scale factor behaves as 
\begin{eqnarray}
a=
\left\{
\begin{array}{ll}
 \left( \frac{\eta_{\rm rh}}{\eta_0}\right) \left( \frac{\eta_{\rm eq}}{\eta_0} \right)
\left(\frac{\eta}{\eta_{\rm rh}}\right)^2&\eta\leq\eta_{\rm rh}\\
 \left (\frac{\eta_{\rm eq}}{\eta_{0}}\right)^2\left(\frac{\eta}{\eta_{\rm eq}} \right)
 &\eta_{\rm rh}\leq\eta\leq\eta_{\rm eq}\\
 \left (\frac{\eta}{\eta_{0}}\right)^2&\eta_{\rm eq}\leq\eta \\
\end{array}
\right.
\end{eqnarray}

Let $\eta_{\rm cross}$ 
($\eta_{\rm rh}<\eta_{\rm cross}<\eta_{\rm eq}$) be the conformal 
time at which the fluctuation with the wavenumber $k$ re-enters the
horizon. Then the spectrum at matter-radiation equality 
($\eta=\eta_{\rm eq}$) is given by 
\begin{eqnarray}
{\cal P}_B(\eta_{\rm eq})=\Bigl( \frac{a_{\rm cross}}{a_{\rm eq}}\Bigr)^4
{\cal P}_B(\eta_{\rm cross}),
\end{eqnarray}
where 
\begin{eqnarray}
{\cal P}_B(\eta_{\rm cross})=\Bigl(\frac{\eta_{\rm cross}}{\eta_{\rm rh}} \Bigr)^2
\Bigl( \frac{a_{\rm rh}}{a_{\rm cross}}\Bigr)^4
{\cal P}_B(\eta_{\rm rh}).
\end{eqnarray}
Setting $z=3$ and $q=2$ (thus $n=4$), ${\cal P}_B(\eta_{\rm rh})$ can be
computed from Eq. (\ref{sp}) as 
\begin{eqnarray}
{\cal P}_B(\eta_{\rm rh})=
 (\alpha M)^{2/3}k^4\frac{\eta_{\rm rh}^2}{a_{\rm rh}^4},
\end{eqnarray}
where $\alpha:=\eta_{\rm eq}/(\eta_{\rm rh}\eta_0^2)$ so that
$a=\alpha\eta^2$ for $\eta\leq\eta_{\rm rh}$.

Since $\eta_{\rm out}<\eta_{\rm rh}$, the horizon re-entry occurs in the
IR regime, where the sound horizon and the Hubble horizon agree. This
implies that $\eta_{\rm cross}\sim k^{-1}$. Thus, after some short
calculations we obtain 
\begin{eqnarray}
\mathcal{P}_{B}(\eta_{\rm eq})  \simeq k^2M^{2/3}a_{\rm eq}^{-7/2}H^{1/3}_{\rm rh}\eta_0^{-1},
\end{eqnarray}
where we have used the relations 
$\eta_{\rm eq} = a_{\rm eq}^{1/2}\eta_0$ and 
$\eta_{\rm rh}\simeq H_{\rm rh}^{-1/2}a_{\rm eq}^{-1/4}\eta_0^{1/2}$.

Now we can evaluate the order of magnitude of the magnetic field at
$\eta=\eta_{\rm eq}$ as 
\begin{eqnarray}
B(\eta_{\rm eq}) & \simeq & \sqrt{\mathcal{P}_B(\eta_{\rm eq})} \nonumber \\
& \simeq & 
 10^{-27}
\Bigl(\frac{k}{1 {\rm Mpc}^{-1}}\Bigr) \Bigl(\frac{M}{10^{-3}M_{\rm pl}} \Bigr)^{1/3}
\Bigl(\frac{a_{\rm eq}}{10^{-3}}\Bigr)^{-7/4} \nonumber \\
& & \times \Bigl(\frac{\eta_0}{13{\rm Gyr}}\Bigr)^{-1/2}
\Bigl(\frac{H_{\rm rh}}{10^{13}{\rm GeV}}\Bigr)^{1/6} {\rm Gauss}.
\label{eqn:Beq}
\end{eqnarray}
At galactic scales, the primordial amplitude (\ref{eqn:Beq}) can be
amplified to the observed amplitude of galactic magnetic fields by the
dynamo mechanism, following the argument given in
Ref.~\cite{Davis:1999bt}.

On the other hand, if there is no amplification mechanism then the
equal-time amplitude (\ref{eqn:Beq}) would correspond to about
$10^{-33}$Gauss or lower for scales of $1$Mpc or longer at present
time. This means that intercluster magnetic field predicted by our
mechanism is too weak to be observed directly or
indirectly~\cite{Kristiansen:2008tx}. 
As for cluster magnetic field, as stated in introduction, there is no
consensus about how observed amplitudes could be related to primordial
ones. It is, however, possible that the primordial magnetic fields
generated by our mechanism at galactic scales could be amplified by the
galactic dynamo and then spread over cluster scales~\cite{ICM1,ICM2}.

Finally we comment on the constraint from the Big Bang Nucleosyntheis (BBN). 
Since the abundance of the light elements is observed precisely, BBN gives 
the upper limit on the strength of the magnetic fields. 
The limits on the homogeneous magnetic fields on the BBN-horizon size ($\sim 10^{-4}$Mpc) 
are less than $10^{-6}$ Gauss in terms of today's values \cite{Grasso:1996kk}. 
The magnetic field generated in our current model is $B\sim 10^{-29}$ Gauss on the 
BBN-horizon scale in terms of today's values. Thus we see that it is consistent with 
the BBN constraint. 

\section{Summary}
\label{sec:summary}

We have presented a new mechanism for generation of large-scale magnetic
field, based on the power-counting renormalizable theory of gravitation
recently proposed by Ho\v{r}ava. Contrary to the usual case in general
relativity, the $U(1)$ gauge symmetry of a Maxwell action in this theory
permits terms breaking conformal invariance in the
ultraviolet. Moreover, for high frequency modes, the anisotropic scaling
intrinsic to the theory inevitably makes the sound horizon far outside
the Hubble horizon. Consequently, non-inflationary cosmic expansion in
the early universe naturally generates super-horizon quantum
fluctuations of the magnetic field. Specializing our consideration to
the case with the dynamical critical exponent $z=3$, we have shown an
explicit set of parameters for which (i) the amplitude of generated
magnetic field is large enough as a seed for the dynamo mechanism; (ii)
backreaction to the cosmic expansion is small enough; and (iii) the
high-energy dispersion relation is consistent with the most recent
observational limits from MAGIC and FERMI.

As stated in Sec.~\ref{sec:intro}, the value of $z$ ($\geq 3$) in the UV
is a part of the definition of a theory, provided that the theory is
renormalizable. In the present paper, we have restricted our
consideration to the simplest case where $z$ in the UV is $3$. However,
in principle the mechanism presented in this paper works for any values 
of $z$ in the UV. For general $z$, radiation energy density scales as
$\rho\propto a^{-(3+z)}$. This means that a radiation dominated epoch of
the universe has a power-law expansion $a\propto t^p$ with
$p=2/(3+z)$. This expansion law satisfies the condition
(\ref{freeze-out}), or $p>1/z$, if $z>3$. For this reason, if we
consider a version of the Ho\v{r}ava-Lifshitz theory with $z>3$ then
magnetic fields can be generated during a radiation dominated
epoch. Further investigation of the mechanism with general $z$ is
certainly worthwhile.

\section*{Acknowledgements}

S. Maeda is supported by the Grant-in-Aid for the Global COE Program
"The Next Generation of Physics, Spun from Universality and Emergence"
from the Ministry of Education, Culture, Sports, Science and Technology
(MEXT) of Japan. 
S. Mukohyama is supported in part by MEXT through a Grant-in-Aid for
Young Scientists (B) No.~17740134 and through WPI Initiative, by JSPS
through a Grant-in-Aid for Creative Scientific Research No.~19GS0219,
and by the Mitsubishi Foundation. 
T. Shiromizu is partially supported by Grant-Aid for Scientific Research
from Ministry of Education, Science, Sports and Culture of Japan
(Nos.~20540258 and 19GS0219), the Japan-U.K. and Japan-India Research
Cooperative Programs.

\appendix
\section{Quantitative analysis of power spectrum}

Here we will carefully compute the power spectrum of the generated magnetic field. 
To do so we will take a gauge-fixing as usual(For example, see Ref. \cite{Horava:2008jf}). 
Without loss of generality, we can choose the gauge of $A_0=0$. Then we may 
want to take the transverse gauge of $\partial_i A_i=0$. For example, we consider 
the UV action of $z=3$. In this case one of the field equations is  
\begin{eqnarray}
\partial_t \partial_i A_i-\Delta A_0=0,
\end{eqnarray}
where $\Delta := \delta^{ij}\partial_i \partial_j$. From this we can easily see that $\partial_i A_i=0$ holds for all $t$ if we impose 
$\partial_i A_i=0$ at an initial time. 
In addition, we can see that the gauge condition is consistent with the remaining field equations. 
Thus we adopt the gauge of $A_0=\partial_i A_i=0$ hereafter.


In the UV limit the action with the critical exponent $z$ is approximately given by 
\begin{eqnarray}
 S_{UV}=
 \frac{1}{2}\int d\eta~d^3\vec{x}
   \Biggl[(\partial_\eta A_i)^2 + \frac{(-1)^{z+1}}{(aM)^{2z-2}}
          A_i \Delta^zA_i
\Biggr].
\end{eqnarray}
From now on we follow the conventional second quantization. First we expand the vector 
perturbation as 
\begin{eqnarray}
A_i = \int d^3\vec{k}\sum_{\sigma=1,2}\left(
b_{\vec{k},\sigma}u_{\vec{k}}\epsilon_i(\vec{k},\sigma) 
+ b_{\vec{k},\sigma}^\dagger u_{\vec{k}}^\ast\epsilon^\ast_i(\vec{k},\sigma)
\right),
\end{eqnarray}
where $\epsilon_i(\vec{k},\sigma)~(\sigma=1,2)$ is the 
orthonormal transverse polarization vector and the operators 
$b_{\vec{k}}$ and $b_{\vec{k}}^\dagger$ satisfy the following commutators
\begin{eqnarray}
& & \left[b_{\vec{k},\sigma},b_{\vec{k}',\sigma'}^\dagger\right]
=(2\pi)^3\delta_{\sigma\sigma'}\delta^{(3)}(\vec{k}-\vec{k}'),\nonumber \\
& & \left[b_{\vec{k},\sigma},b_{\vec{k}',\sigma'}\right]
=\left[b_{\vec{k},\sigma}^\dagger,b_{\vec{k}',\sigma'}^\dagger\right]=0.
\end{eqnarray}
The vacuum is defined by 
\begin{eqnarray}
b_{\vec{k},\sigma}|0\rangle = 0 ~~~{\rm for}~\vec{k}.
\end{eqnarray}
The mode functions follows the Klein-Gordon normalization as usual: 
\begin{eqnarray}
(u_{\vec{k}},u_{\vec{k}'}) & \equiv & -i\int d^3\vec{x}~
\left(u_{\vec{k}}\partial_\eta u_{\vec{k}'}^\ast 
- u_{\vec{k}'}^\ast\partial_\eta u_{\vec{k}}\right) \nonumber \\
& = & 
\frac{1}{(2\pi)^3}\delta^{(3)}(\vec{k}-\vec{k}').
\label{inner}
\end{eqnarray}
Then the equation for the mode function becomes 
\begin{eqnarray}
u''_{\vec{k}} + \frac{(-1)^z}{(aM)^{2z-2}}\Delta^z u_{\vec{k}}=0,
\end{eqnarray}
where the prime stands for the  derivative with respect to 
the conformal time. Introducing $\chi_{\vec{k}}$ as 
\begin{eqnarray}
u_{\vec{k}}(k,\eta)
=\frac{e^{i\vec{k}\cdot\vec{x}}}{(2\pi)^3}\chi_{\vec{k}}(\vec{k},\eta),
\label{modefunc}
\end{eqnarray}
the equation for $\chi_{\vec{k}}$ becomes 
\begin{eqnarray}
\chi''_{\vec{k}} + \frac{k^{2z}}{(aM)^{2z-2}} \chi_{\vec{k}}=0.
\end{eqnarray}
Here we define that the physical frequency is 
$\omega:=k^z/(a^zM^{z-1})$.


Hereafter, we assume that the scale factor 
has the power law, $a=\alpha \eta^q$, with 
$q=p/(1-p)>1/(z-1)$ which comes from the condition of 
Eq. (\ref{freeze-out}). Then the equation for $\chi_{\vec{k}}$ 
becomes 
\begin{eqnarray}
\chi''_{\vec{k}} + \beta_{k}\eta^{-(2z-2)q}\chi_{\vec{k}}=0,
\end{eqnarray}
where $\beta_{k}:=k^{2z}/(\alpha M)^{2z-2}$. 
The solution is given by
\begin{eqnarray}
\chi_{\vec{k}}
& = & 
C_1\sqrt{\eta}
H^{(1)}_\nu\left(-2\nu\sqrt{\beta_k}\eta^{1/(2\nu)}\right) \nonumber \\
& & +
C_2\sqrt{\eta}
H^{(2)}_\nu\left(-2\nu\sqrt{\beta_k}\eta^{1/(2\nu)}\right),
\end{eqnarray}
where $\nu:=-1/2(qz-q-1)$ and $H^{(n)}_\nu$ is the $\nu$th-order 
Hankel functions. From the normalization and choosing the 
mode function to be the positive-frequency mode in Minkowski 
spacetime at short-wavelength limit, $C_1$ and $C_2$ are 
fixed and 
then 
\begin{eqnarray}
u(\vec{k},\eta)
& = & 
\frac{e^{i\vec{k}\cdot\vec{x}}}{(2\pi)^3}
\sqrt{-\frac{\pi\nu\eta}{2}}
H^{(1)}_\nu
\left(-2\nu\sqrt{\beta_k}\eta^{1/(2\nu)}\right) 
e^{i\pi\frac{2\nu+1}{4}}. \nonumber \\
\end{eqnarray}

Now we can calculate the power spectrum of 
$A_i$
which is defined as 
\begin{eqnarray}
\langle 0|A_{i,\vec{k}} A_{j,\vec{k}'} |0\rangle 
\equiv 
(2\pi)^3\delta_{ij}\delta^{(3)}(\vec{k}+\vec{k}')
\frac{2\pi^2}{k^3}\mathcal{P}_{A_i},
\end{eqnarray} 
where 
\begin{eqnarray}
A_{i,\vec{k}} 
= \int d^3\vec{x}~e^{i\vec{k}\cdot\vec{x}} A_i(\vec{x}).
\end{eqnarray}
Then the power spectrum for $A_i$ will be  
\begin{eqnarray}
\mathcal{P}_{A_i} 
&=& \frac{k^3}{2\pi^2}\left|(2\pi)^3 
    u_{\vec{k}}\right|^2 \nonumber \\
& = & \frac{k^3}{2\pi^2}
\left(-\frac{\pi\nu\eta}{2}\right)
\left|H^{(1)}_\nu
\left(-2\nu\sqrt{\beta_k}\eta^{1/(2\nu)}\right)
\right|^2.
\end{eqnarray}
Therefore, the power spectrum of the magnetic field is
\begin{equation}
 {\cal P}_{B} \simeq \frac{k^2}{a^4}{\cal P}_{A_i}. 
\end{equation}

Now we estimate $\mathcal{P}_{A_i}$ and ${\cal P}_{B}$ in 
$\eta \to \infty$ when the fluctuations freeze out
($H \gg \omega $). 
The mode function will be approximated by 
\begin{eqnarray}
u(\vec{k},\eta)
&=&
\frac{e^{i\vec{k}\cdot\vec{x}}}{(2\pi)^3}
\sqrt{\frac{-\nu}{2\pi}}\eta\Gamma(-\nu)
\left(-\nu\sqrt{\beta_k}\right)^\nu
e^{i\pi\frac{2\nu+1}{4}}. \nonumber \\
& & 
\end{eqnarray}
Then the power spectrum is obtained as
\begin{eqnarray}
\mathcal{P}_{A_i}|_{H\gg \omega} = 
\frac{k^3(-\nu)^{2\nu+1}}{4\pi^3}
\left(\eta\Gamma(-\nu)\right)^2
(\beta_k)^{\nu}.
\label{PA}
\end{eqnarray}
Finally ${\cal P}_{B}|_{H\gg \omega}$ becomes 
\begin{eqnarray}
{\cal P}_{B}|_{H\gg \omega} 
& \simeq & 
\frac{k^5(-\nu)^{2\nu+1}}{4\pi^3a^4}
\left(\eta\Gamma(-\nu)\right)^2
(\beta_k)^{\nu} \nonumber \\
& \simeq & (\alpha M)^{(z-1)/(qz-q-1)}k^n\frac{\eta^2}{a^4},
\label{mag1}
\end{eqnarray}
where $n=5-z/(qz-q-1)$. 
This result agrees with that of Eq. (\ref{sp}) using the scaling law argument 
in the text.


\end{document}